\documentclass[twocolumn, superscriptaddress, prb, showpacs]{revtex4-1}

\usepackage{graphicx}
\usepackage{bm}
\usepackage{amsmath}
\usepackage{color}

\graphicspath{{./figures/}}
\newcommand{\br}{\mathbf{r}}
\newcommand{\bq}{\mathbf{q}}
\newcommand{\bqm}{\mathbf{q'}}
\newcommand{\diff}{\mathrm{d}}
\newcommand{\eps}{\epsilon}
\newcommand{\im}{\mathrm{i}}
\newcommand{\Enl}{E_\text{c}^\nl}
\newcommand{\nl}{\text{nl}}
\newcommand{\sloc}{\text{sl}}

\begin{document}


\title{van der Waals density functional with corrected $C_6$ coefficients}

\author{K. Berland}
\email[Email: ]{kristian.berland@nmbu.no}
\affiliation{Faculty of Science and Technology, Norwegian University of
Life Sciences, Norway.}
\affiliation{Centre for Materials Science and Nanotechnology, University
of Oslo, Norway.}

\author{D. Chakraborty}
\affiliation{Department of Physics, Wake Forest University,
Winston-Salem, NC 27109, USA.}
\affiliation{Center for Functional Materials, Wake Forest University,
Winston-Salem, NC 27109, USA.}

\author{T. Thonhauser}
\email[E-mail: ]{thonhauser@wfu.edu}
\affiliation{Department of Physics, Wake Forest University,
Winston-Salem, NC 27109, USA.}
\affiliation{Center for Functional Materials, Wake Forest University,
Winston-Salem, NC 27109, USA.}

\date{\today}

\begin{abstract}
The non-local van der Waals density functional (vdW-DF) has had
tremendous success since its inception in 2004 due to its
constraint-based formalism that is rigorously derived from a many-body
starting point. However, while vdW-DF can describe binding energies and
structures for van der Waals complexes and mixed systems with good
accuracy, one long-standing criticism---also since its inception---has
been that the $C_6$ coefficients that derive from the vdW-DF framework
are largely inaccurate and can be wrong by more than a factor of two. It
has long been thought that this failure to describe the $C_6$
coefficients is a conceptual flaw of the underlying plasmon framework
used to derive vdW-DF. We prove here that this is not the case and that
accurate $C_6$ coefficient can be obtained without sacrificing the
accuracy at binding separations from a modified framework that is fully
consistent with the constraints and design philosophy of the original
vdW-DF formulation. Our design exploits a degree of freedom in the
plasmon-dispersion model $\omega_{\bq}$, modifying the strength of the
long-range van der Waals interaction and the cross-over from long to
short separations, with additional parameters tuned 
to reference systems. Testing the new formulation for a range of different
systems, we not only confirm the greatly improved description of $C_6$
coefficients, but we also find excellent performance for molecular
dimers and other systems. The importance of this development is not
necessarily that particular aspects such as $C_6$ coefficients or
binding energies are improved, but rather that our finding opens the
door for further conceptual developments of an entirely unexplored
direction within the exact same constrained-based non-local framework
that made vdW-DF so successful in the first place.
\end{abstract}


\maketitle

\section{Introduction}\label{sec:introduction}
Materials in which van der Waals interactions, i.e.\ London dispersion
forces, play a crucial role for cohesion and binding properties now
stand at the forefront of a number of major scientific and technological
advances. Examples include gas storage and filtering in supermolecular
complexes and porous materials,\cite{Tan_2016:trapping_gases,
Wang_2018:topologically_guided, Li_2017:capture_organic} organic
electronic and optoelectronic
applications,\cite{Diemer_2017:influence_isomer,
Gomez-Bombarelli_2016:design_efficient} and
pharmaceutical,\cite{aspirin}
ferroelectric,\cite{Lee_2012:structure_energetics,
Shoji2018:Ferroelectric} and photovoltaic molecular
crystals.\cite{molecular_crystals:review,Rangel2016} Common to several
of these developments is the increasing role of first-principle
electronic-structure calculations at the density functional theory (DFT)
level---serving not only to gain insight into their functionality, but
also to predict new materials and functionality prior to experimental
synthetization.\cite{Gomez-Bombarelli_2016:design_efficient} As the
ability to design and analyze materials often hinges on the ability to
accurately predict energetic and structural properties, over the years
great effort has been made to improve DFT. In this regard, a major
development was the inclusion of van der Waals interactions by various
means during the previous decade.\cite{Grimme_2004:accurate_description,
Grimme_2007:density_functional, Grimme:review, Grimme:review,
Dion_2004:van_waals, Langreth_2009:density_functional,
Berland_2015:van_waals, Tkatchenko_2009:accurate_molecular,
Tkatchenko_2012:accurate_efficient, MBD:update, wavelikeMBD:2016,
Vydrov_2009:improving_accuracy, Vydrov_2009:nonlocal_van,
Vydrov_2010:dispersion_interactions, Vydrov_2010:nonlocal_van} However,
unfortunately, current methods to treat van der Waals interactions still
have not achieved the same level of accuracy and reliability for such
non-covalently bonded systems as what is now typical for
covalently-bonded ones. 

Amongst the various methods to capture van der Waals interactions within
DFT, the vdW-DF method developed by Langreth, Lundqvist, and
coworkers\cite{Dion_2004:van_waals, Langreth_2009:density_functional,
Lee_2010:higher-accuracy_van, Berland_2015:van_waals,
Thonhauser_2007:van_waals, Schroder_2017:vdw-df_family,
Thonhauser_2015:spin_signature, Berland_2017:asssement_hybrid} stands
out in that it is a true density functional, i.e.\ it can be evaluated
from knowledge of the density alone, and employs a non-local correlation
functional $\Enl[n]$ to account for dispersion forces. The tremendous
success of vdW-DF is rooted in its plasmon dispersion formalism that is
rigorously derived from a many-body starting point and adheres to a
number of exact physical constraints.\cite{Dion_2004:van_waals,hybesc14} This
non-local correlation functional has since become the cornerstone of
several higher-level improvements that e.g.\ adjust the exchange
functional that is being used in conjunction with
$\Enl[n]$.\cite{Cooper_2010:van_waals, Klimes_2010:chemical_accuracy,
Klimes_2011:van_waals, Hamada_2014:van_waals} However, despite its
success and widespread use, further conceptual development of vdW-DF has
come to a halt---although around the turn of the last decade vdW-DF
inspired the related and well-crafted functionals by Vydrov and Voorhis
(VV),\cite{Vydrov_2009:improving_accuracy, Vydrov_2009:nonlocal_van,
Vydrov_2010:dispersion_interactions, Vydrov_2010:nonlocal_van} its
fundamental framework has not changed since 2004. Furthermore, one
common criticism has also plagued vdW-DF for almost the same time span,
i.e.\ the often poor $C_6$ coefficients that derive from it,
\cite{Vydrov_2010:dispersion_interactions, Woods_2016:materials_perspective,Schroder_2017:vdw-df_family}\footnote{Private communications.}
which has been thought to be a conceptual flaw in the underlying framework. The
poor $C_6$ coefficients in vdW-DF themselves bare little to no effect
on the vdW-DF performance for binding energies and structures, but they
are a formal shortcoming nonetheless. We prove here that this is not a
conceptual flaw of the framework and that a modification of the
underlying plasmon dispersion model corrects the $C_6$ coefficients and
opens the door for utilizing an entirely unexplored degree of freedom
that may be used for further improvements---all while adhering to the
same physical constrains that made the original vdW-DF formalism so
transferable and successful.

To understand which changes are necessary inside vdW-DF, we
review in the next section the underlying framework but defer the reader
to in-depth discussions elsewhere.\cite{Dion_2004:van_waals,
Berland_2015:van_waals, Thonhauser_2007:van_waals,
Schroder_2017:vdw-df_family, Thonhauser_2015:spin_signature} The
non-local correlation functional $\Enl[n]$ in vdW-DF can be derived as a
systematic expansion of the adiabatic connection formula
(ACF)\cite{gulu76, lape75, Langreth_1977:exchange-correlation_energy} in
terms of an effective plasmon propagator $S$, which has poles for real
frequencies at the effective plasmon frequency $\omega_{\bq}$, where
$\bq$ is the momentum of the plasmon. A number of exact physical
constraints on $S$ and $\omega_\bq$ explain the transferability and
success of vdW-DF in describing diverse classes of
materials.\cite{Langreth_2009:density_functional,
Berland_2014:van_waals, Schroder_2017:vdw-df_family, Berland_2015:van_waals} However, the $C_6$
coefficients in vdW-DF arise exclusively from the $\bq \to 0$ limit of
the plasmon dispersion $\omega_\bq$, which is not constrained in vdW-DF
and arises merely as a byproduct of the particular parametrization of
$\omega_\bq$. As a result, vdW-DF typically exhibits inferior $C_6$
coefficients compared to other methods.

Here, we modify the vdW-DF framework, fully consistent with the original
constraints and design philosophy, but with a new and more flexible parameterization of
the plasmon dispersion $\omega_\bq$. This flexibility is exploited to
provide accurate $C_6$ values resulting in a mean absolute relative
deviation (MARD) of the $C_6$ coefficients of 11\% compared to 20\%
for the original formulation, which we refer to here as
vdW-DF1,\cite{Dion_2004:van_waals} and 56\% for
vdW-DF2.\cite{Lee_2010:higher-accuracy_van}
In addition---although not
the primary purpose of this paper---the form of $\omega_\bq$ is also tuned
to provide noticeable improvements for binding energies of several molecular dimer systems,
making the functional a contender for applications to this class of systems.
Finally, the new formulation is also trivial to implement in compute codes with existing vdW-DF
implementations.

\section{Theory }\label{sec:theory}

The non-local correlation in vdW-DF is given by the second-order
expansion of the ACF in terms of a plasmon propagator $S$, describing
virtual charge-density fluctuations of the electron gas, as follows
\begin{align}
E_c^{\nl}[n] =&\int_{0}^{\infty}\frac{\diff u}{4\pi}\int \frac{\diff^3 \bq}{(2\pi)^3}\,
\frac{\diff^3 \bq' }{(2\pi)^3 } \nonumber \\ &
	\times\left[ 1 - (\hat{\bq}\cdot \hat{\bq}')^2 \right] S_{\bq,\bqm}(\im u)\,
	S_{\bqm,\bq}(\im u)\;,
\label{eq:secondWwithSb}
\end{align}
where $S$ is given by
\begin{align}
	S_{\bq,\bq'}(\im u) = & \;\frac{1}{2} \, \big[\tilde{S}_{\bq,\bq'}(\im u) +
\tilde{S}_{\bq',\bq}(\im u)\big]\label{equ:S} \\
\tilde{S}_{\bq,\bq'}(\im u) = & \int \diff^3\br \; \frac{4\pi\, n(\br)\; e^{-\im(\bq-\bq')\cdot\br}}
{\big(\im u+\omega_\bq(\br)\big)\big(-\im u+\omega_{\bq'}(\br)\big)} \;.
\label{equ:Stilde}
\end{align}
Here, $\omega=\im u$ is the imaginary frequency, $n(\br)$ is the electron density,
and $4\pi n(\br)$ is the square of the plasmon frequency. This
particular form of $S$ as been chosen because it fulfills four exact
physical constraints,\cite{Dion_2004:van_waals} making vdW-DF such a
powerful and transferable tool. In our modification of the method, we
retain the same model of $S$ as in earlier versions, but update the
plasmon dispersion $\omega_\bq(\br) =
q^2/\big[2\,h\big(q/q_0(\br)\big)\big]$ by modifying the dimensionless
switching function $h\big(q/q_0(\br)\big)$ which controls the relative
strength of the density response at different length scales. The
function $q_0(\br)$ parameterizes the local response of the electron gas
and is determined by the requirement that the first-order term in $S$ in
the ACF expansion reproduces a GGA-type XC functional.\cite{hybesc14,
Berland_2015:van_waals, Schroder_2017:vdw-df_family} This XC functional,
which is generally not the same as in the total energy functional, is
named the {\it internal} functional $\eps^{\rm int}_{\rm
xc}$.\cite{Dion_2004:van_waals, Berland_2014:exchange_functional,
Berland_2015:van_waals} $q_0(\br)$ is related to the
exchange-correlation per particle through this first-order expression in
$S$ as follows
\begin{align}
\eps^{\rm int}_{\rm xc}(\br) &= \pi \int \frac{\diff^3 \bq}{(2\pi)^3} \left[ \frac{1}{\omega_\bq(\br)} - \frac{2}{q^2} \right]  \nonumber \\
&= 2 \pi \int \frac{\diff^3 \bq}{(2\pi)^3} \frac{1}{q^2} \left[ h\big(q/q_0(\br)\big) -1 \right]\nonumber\\
&= -\frac{1}{\pi}\,q_0(\br)\int_0^\infty \diff y \, [1 - h(y) ] \,. \label{eq:eq1}
\end{align}
If we constrain the remaining integral to be
\begin{align}
\int_0^\infty \diff y \, [1 - h(y) ] = \frac{3}{4} \label{eq:cond}\;,
\end{align}
then $q_0(\br)$ can conveniently be expressed as a modulated Fermi wave
vector $k_{\rm F}^3(\br)=3\pi^2n(\br)$, i.e.\
$q_0(\br)=-(4\pi/3)\,\eps^{\rm int}_{\rm xc}(\br)=\big(\eps^{\rm
int}_{\rm xc}(\br)/\eps_{\rm x}^{\rm LDA}(\br)\big)\,k_{\rm F}(\br)$.

The vdW-DF framework dictates three constraints on $h(y)$, but also
leaves considerable freedom. First, the integral over $h(y)$ should be
constrained by Eq.~(\ref{eq:cond}). Second, a quadratic small-$y$ limit
in $h(y) = \gamma y^2 + \dots$ (where $\gamma$ is a constant) ensures
the appropriate $C_6/r^6$ long-range limit. 
In fact, Hyldgaard~et~al.\cite{hybesc14} pointed out that $h(0) = 0$
corresponds to charge conservation of the spherical XC hole model of the internal functional, which is given by a Fourier transformation of $h\big(q/q_0(\br)\big)$. 
Third, $h(y)$ is required to
increase monotonically to a large-$y$ limit of
$\lim_{y\to\infty}h(y)=1$, corresponding to $\omega_\bq \to q^2/2$ in
the limit of large $q$. This allows $S$ to cancel the self-interaction
divergence, i.e.\ the $2/q^2$ term in Eq.~(\ref{eq:eq1}), in the ACF
formula.\cite{Dion_2004:van_waals,Berland_2014:van_waals} The standard switching
function in vdW-DF1 and vdW-DF2 was chosen within these constraints as
\begin{equation}
h_{\rm std}(y)= 1-\exp(-\gamma y^2)\;,
\end{equation}
where the value of $\gamma=\gamma_{\rm std}=4\pi/9\approx1.3963$ is
determined by Eq.~(\ref{eq:cond}). Note that $h_{\rm std}(y)$
includes only one parameter, which is fully constrained.

Dispersion forces in vdW-DF can be viewed as
arising from a coupling of semi-local XC holes of separated
bodies.\cite{hybesc14} For two such bodies $A$ and $B$ far apart, the
$C_6$ coefficient is given by
\begin{align}
C_6 =  \frac{3}{\pi}\int_0^\infty \diff u \; \alpha_A(\im u)\,\alpha_B(\im u)\;,\label{eq:C6}
\end{align}
with the far-field polarizability of body $A$ (and likewise for $B$) given by
\begin{align}\label{eq:alpha}
\alpha_A(\im u) &= \int_A \diff ^3\br\, \frac{n(\br)}{\omega_0^2(\br) + u^2}\,,
\end{align}
where $\omega_0(\br) = \lim_{\bq\to 0}\omega_{\bq}(\br)=q_0(\br)^2/2
\gamma$ for any $h$ function with the required small-$y$ limit from the
second constraint. It is very interesting to see that the first
constraint is a condition for the integral over all values of $y$, but
the asymptotic limit that determines the $C_6$ coefficients is
determined exclusively by the small-$y$ limit in the second constraint.
Crucially, there is significant residual freedom in the form of possible
$h(y)$ functions that still fulfill all three constraints. Specifically,
from the second constraint it follows that $\gamma = {\rm lim}_{y
\rightarrow 0 } h(y)/y^2$, which could take almost any number. Moreover,
as $C_6 \propto \gamma^3$, controlling the value of $\gamma$ is
paramount to securing accurate $C_6$ values.

\begin{figure}
\includegraphics[width=8cm]{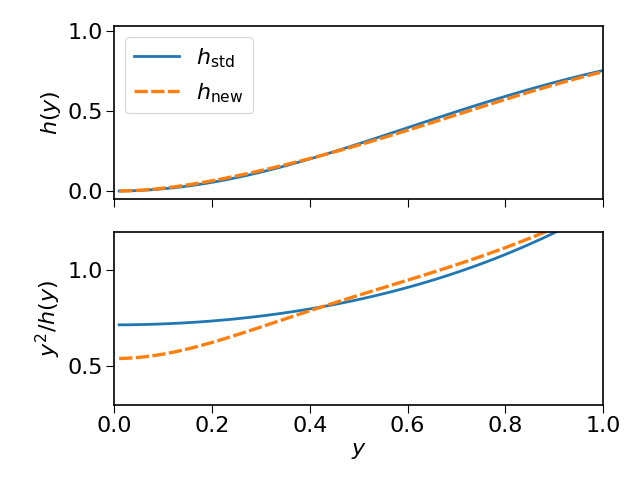}
\caption{Comparison of the standard switching function $h_{\rm std}(y)$
used in vdW-DF1 and vdW-DF2 and the new switching function $h_{\rm
new}(y)$. The upper panel shows $h(y)$, while the lower panel shows
$y^2/h(y)$ which is proportional to the plasmon pole $\omega_\bq$.
Somewhat surprisingly, the overall shape of both functions is strikingly
similar, which is related to the fact that they obey the same underlying
constraints. However, in the $y<0.4$ region there are significant
differences, most visible in the bottom panel.\label{fig:new_h}}
\end{figure}

At the heart of our modification is a new switching function that still
fulfills all three constraints, but that is also crafted to improve the
description of $C_6$ coefficients. In particular, we propose
\begin{align}
h_{\rm new}(y) & =  1 - \left(1 + \frac{  (\alpha - \gamma) y^2 +  A(\alpha,\gamma, \beta) y^4}{1 + A(\alpha,\gamma, \beta) y^2} \right) \exp(-\alpha y^2),
\end{align}
where $A(\alpha, \gamma, \beta) = \big( \beta + \alpha(\alpha/2 - \gamma)
\big)/\left( 1 + \gamma -\alpha \right)$. This three-parameter function
$h_{\rm new}(y)$ provides significantly more freedom than $h_{\rm
std}(y)$. We get the required small-$y$ limit of $h_{\rm new}(y) =
\gamma y^2 - \beta y^4 + \dots$, or equivalently 
\begin{align}
\frac{y^2}{h_{\rm new}(y)} = \frac{1}{\gamma} + \frac{\beta}{\gamma^2} y^2 + \mathcal{O}(y^4)\,,
\end{align}
which shows that---while $\gamma$ controls the magnitude of the
asymptote and $C_6$ coefficients---$\beta/\gamma^2$ is the leading order term
determining how van der Waals interactions are reduced at shorter
separations. We found the optimal values to be $\alpha = 2.01059$,
$\beta = 8.17471$, and $\gamma=\gamma_{\rm new}=1.84981$, as described
below. The functions $h_{\rm new}(y)$ and $h_{\rm std}(y)$ are plotted
in Fig.~\ref{fig:new_h} for comparison.
The non-local correlation energy in Eq.~(\ref{eq:secondWwithSb}) can also be written
as $E_c^{\rm nl}[n]=\frac{1}{2}\int\diff^3\br\,\diff^3\br'\;n(\br)\,\phi(\br,\br')\,n(\br')$
where the kernel $\phi(\br,\br')$ is fully determined through Eq.~(\ref{eq:secondWwithSb})--(\ref{equ:Stilde}); the two switching functions $h_{\rm new}(y)$ and $h_{\rm std}(y)$ result in two different kernels
$\phi_{\rm new}(\br,\br')$ and $\phi_{\rm std}(\br,\br')$, the difference of which is plotted in Fig.~S1.


The value of $\gamma_{\rm new}=1.84981$ was determined by minimizing the
MARD of a set of 34 closed-shell atoms and small molecules compiled by
Vydrov and Voorhis;\cite{Vydrov_2010:dispersion_interactions}
computational details are provided in Appendix
\ref{sec:computational_details}. For a given internal functional, tuning
the $\gamma$ parameter corresponds to scaling the $C_6$ coefficients by
$\gamma^3$. Note that vdW-DF1 and vdW-DF2 use different
internal functionals, as described in Appendix~\ref{sec:internal}.
If we use the internal functional of vdW-DF2, the optimal scaling
$\left(\gamma_{\rm new}/\gamma_{\rm std} \right)^3= 2.32529$, resulting
in a MARD for the $C_6$ coefficients of
11.13\%. If one instead tried to optimize $C_6$ coefficients with the
vdW-DF1 internal functional, the lowest achievable MARD would be 18.56\%---hence, our
modified version of vdW-DF utilizes the vdW-DF2 internal functional.
The value of $\beta = 8.17471$, with a corresponding $\alpha = 2.01059$,
was determined by minimizing the MARD of the binding energies of the S22
data set with separations optimized along the center-of-mass
coordinates, which results in a MARD value of 5.72\%. In the
optimization, for each value of $\beta$, $\alpha$ was adjusted to
fulfill Eq.~(\ref{eq:cond}).

\begin{figure*}[t!]
\includegraphics[width=16cm]{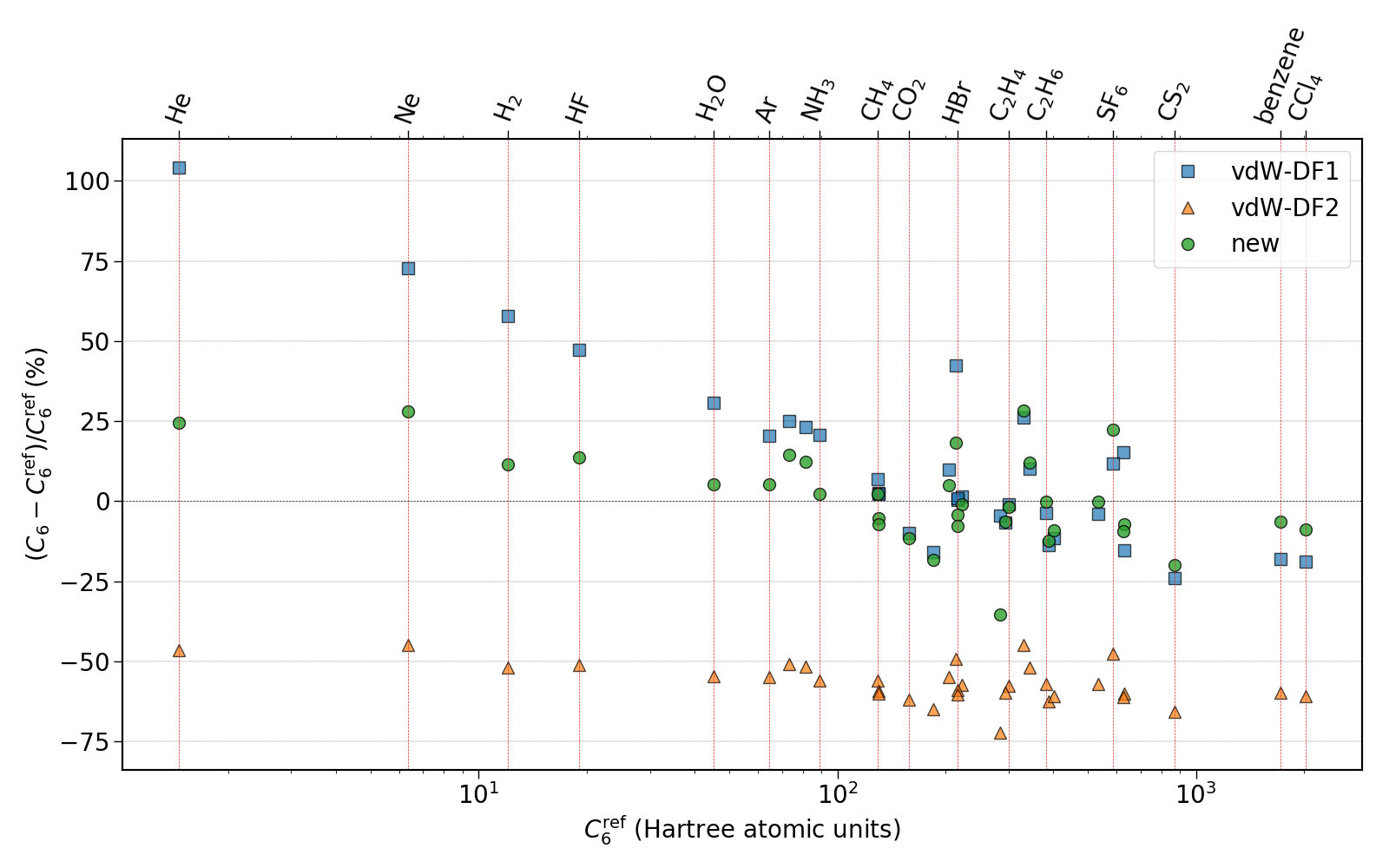}
\caption{\label{fig:C6} Relative deviations of $C_6$ coefficients of
vdW-DF1 (filled squares), vdW-DF2 (triangles), and our new framework
(circles) for a set of 34 systems compiled by Vydrov and
Voorhis, with reference data from a variety of sources.\cite{Vydrov_2010:dispersion_interactions} 
The $C_6$ coefficients are calculated from
Eqs.~(\ref{eq:C6}) and (\ref{eq:alpha}). Selected systems are
labeled along the upper $x$-axis. The corresponding MARD values are:
19.97\% (vdW-DF1), 55.64\% (vdW-DF2), and 11.13\% (new), \ showing the
drastic improvement of $C_6$ values. Explicit values for all $C_6$
coefficients---also including other functionals for comparison---can be
found in Table~\ref{table:C6} in Appendix~\ref{sec:data}.}
\end{figure*}

Finally, the non-local correlation energy from
Eq.~(\ref{eq:secondWwithSb}) has to be combined with a (semi)local
functional to give the entire exchange-correlation energy $E_{\rm xc}[n]$,
\begin{equation}
E_{\rm xc}[n] = E_{\rm xc}^{\sloc}[n] + E_{\rm c}^{\nl}[n]\;.
\end{equation}
Only LDA correlation is included in $E_{\rm xc}^{\sloc}[n]$ as in all standard vdW-DF variants,  whereas for the exchange contribution of $E_{\rm xc}^{\sloc}[n]$ we employ the
B86R functional of Hamada.\cite{Hamada_2014:van_waals} 
Note that our
choice for the $\beta$ parameter is dependent on the exchange functional
employed. Since $\beta$ is optimized for a reference set of systems, the
binding energies are less sensitive to the exchange functional choice than
in standard vdW-DF. Nonetheless, the exchange has a strong impact on
binding separations.\cite{londero11p1805, berland11p1800, behy13,
Berland_2014:van_waals} vdW-DF variants based on soft exchange
functionals, such as C09,\cite{Cooper_2010:van_waals} optB86b,\cite{Klimes_2011:van_waals} and cx13\cite{Berland_2014:exchange_functional} have generally been found to
be more versatile than those based on the harder revPBE\cite{Zhang_1998:comment_generalized} and PW86r\cite{Murray_2009:investigation_exchange} originally employed in vdW-DF1\cite{Dion_2004:van_waals} and vdW-DF2.\cite{Lee_2010:higher-accuracy_van}
The B86R\cite{Hamada_2014:van_waals}
functional of Hamada was chosen because it was constructed for vdW-DF2 and we retain the same internal functional as vdW-DF2. 
vdW-DF2-B86R provides accurate binding separations
for layered and adsorption systems, molecular dimers, and lattice
constants of solids. Moreover, the B86R exchange functional goes as $s^{2/5}$
in the large-$s$ limit, where $s$ is the reduced gradient, which Murray
et al.\cite{Murray_2009:investigation_exchange} argued to be a suitable
choice as it avoids spurious long-range binding effects in the exchange
channel. The cx13 exchange functional used in vdW-DF1-cx was not employed
because this functional was originally constructed for consistency with
the internal exchange functional of vdW-DF1.

\section{Results}\label{sec:results}

The parameters in our new switching function have been chosen to
provide accurate $C_6$ coefficients and a low MARD for the S22 set. We
first present here values for the $C_6$ coefficients of
our new framework, followed by a systematic test of the performance of our new modification to ensure that the improved
$C_6$ coefficients do not come to the detriment of worse performance in
other areas. As such, we consider a range of standard test systems---specifically, the S66 set of molecular dimers relevant for biomolecular systems, the X40 set of halogenated
molecules, and 23 solids. However, more in depth testing, also considering
challenging extended systems such as molecular crystals and surface adsorption
would be necessary if we were to release our new formulation as a general purpose
functional---but that is not the intent and we merely want to show that the
improved $C_6$ coefficients come with reasonable performance in other areas.
To assess strengths and weaknesses of our
modifications, we also applied  a number
of other commonly employed van der Waals functionals to the same systems. Details on our computational approach can be found in Appendix~\ref{sec:computational_details}.
See also Supplementary Information (SI) at for detailed results
for all the individual systems presented in this paper.\footnote{See Supplemental Material [url], which includes Refs.\onlinecite{Schroder_2017:vdw-df_family,Dion_2004:van_waals,Takatani2010:S22dataset_revised,Pavel2011:S66dataset,Pavel2012:X40dataset,Klimes_2011:van_waals}}

\subsection{$C_6$ Coefficients}

The improvement in $C_6$ values are illustrated in Fig.~\ref{fig:C6},
with numerical data provided in Table \ref{table:C6} in
Appendix~\ref{sec:data}, for a set of 34 systems compiled by Vydrov and
Voorhis.\cite{Vydrov_2010:dispersion_interactions} Figure~\ref{fig:C6}
shows that---while vdW-DF2 consistently underestimates $C_6$
coefficients by a factor of approximately 2---their trends are far more
accurately described than in vdW-DF1, which overestimates the $C_6$
coefficients of small molecules. This behavior is the reason for the choice of the
internal functional of vdW-DF2 for our new formulation, allowing for a
smaller MARD with a common scaling factor
$(\gamma_{\rm new}/\gamma_{\rm std})^3$ that corrects the $C_6$
coefficients for a wide range. Overall, we find the expected
improvement of the MARD value for $C_6$ coefficients,
which now drops to 11.13\% for our new formulation compared to
19.97\% for vdW-DF1 and 55.64\% for vdW-DF2.  This value is on
par with the very best of all functionals tested, i.e.\ 10.73\% for rVV10
in Table~\ref{table:C6}, albeit the VV formalism has been derived either
by relaxing some of the original constraints of vdW-DF or, in the case
of the widely employed VV10 functional,\cite{Vydrov_2010:nonlocal_van}
through heuristic rationalization.

\subsection{Molecular Dimers}

\subsubsection{The S22 Set}

\begin{figure}
\includegraphics[width=\columnwidth]{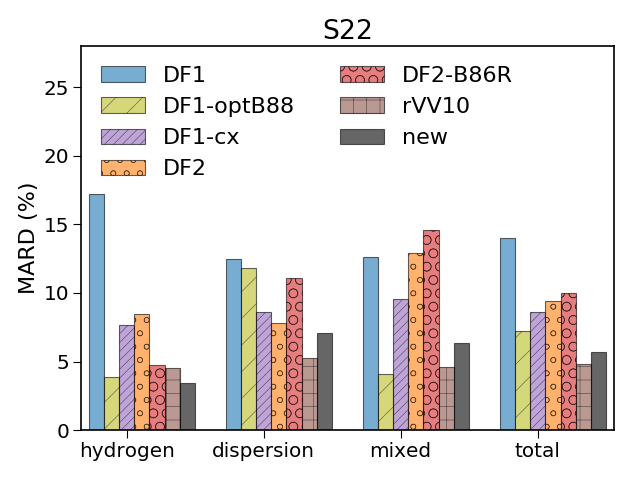}
\caption{\label{fig:S22MARD} MARD of the binding energy for the S22 data
set, split into the characteristic hydrogen- and dispersion-bonded systems
and mixtures of these bonding types.}
\end{figure}

Figure~\ref{fig:S22MARD} shows the MARD of the binding energies of the S22 set of molecular dimers.
More detailed data is available in the Appendix in Table~\ref{table:S22}. 
The fact that vdW-DF1-optB88, rVV10, and our new development has a MARD smaller than $\sim$7\%  is a
result of optimizing either the exchange (in vdW-DF1-optB88) or correlation (in rVV10 and our new development) to this data set. Because of this bias, we will not further analyze
the results for the S22 data set.

\subsubsection{The S66 Set}

\begin{figure}
\includegraphics[width=\columnwidth]{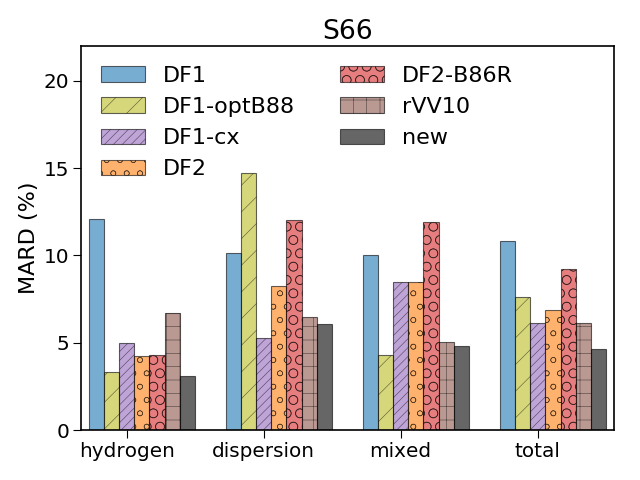}
\caption{\label{fig:S66MARD}MARD of the binding energy for the S66 data
set, split into the characteristic hydrogen- and dispersion-bonded systems
and mixtures of these bonding types.}
\end{figure}

The S66 set of molecular dimers is larger and more diverse than the S22, featuring 
several dimers involving non-aromatic molecules and double-hydrogen bonds. Moreover,  reference binding separations are more accurate.\cite{Pavel2011:S66dataset}

\begin{figure}
\includegraphics[width=\columnwidth]{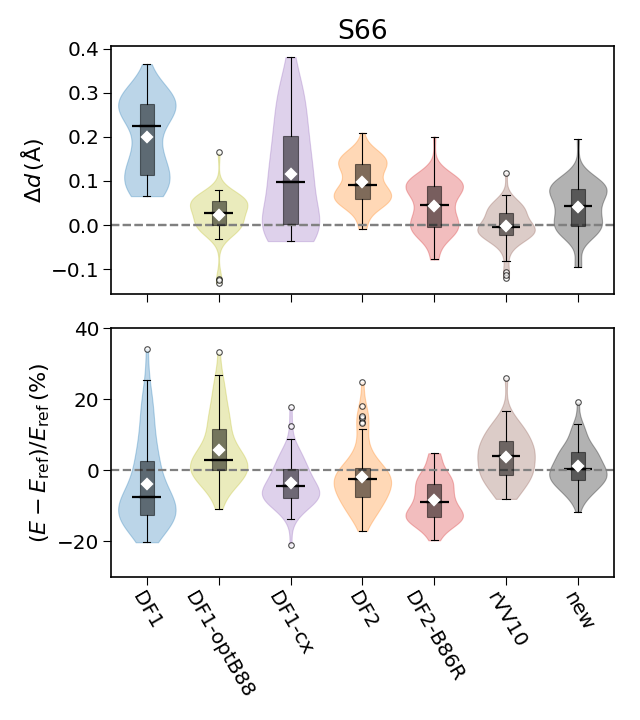}\\
\caption{\label{fig:S66_box} Visualization (violin plot overlaid a box plot) of the deviations from reference data of the different functionals for the S66 data set\label{fig:S66_violon}. The upper panel gives deviations in separations and the lower panel shows the relative deviation of the binding energy.
The violin plots (transparent color) represent the data distribution and are based on a Gaussian kernel density estimation using the Scott's rule\cite{Scott:violin} as implemented in \textsc{matplotlib}. 
In the box plot, the boxes hold 50\% of the data, with equal number of data points above and below the median deviation (full black line). The whiskers indicate the range of data
falling within 1.5$\times$box-length beyond the upper and lower limits of the box. Outliers beyond this range are indicated with circular makers and are identified in the SI.\cite{Note2}
The diamonds mark the mean deviation.}
\end{figure}

Figure~\ref{fig:S66MARD} shows the MARD of the binding energies of the S66 set and Table~\ref{table:S66} provides details on their statistical properties. 
In addition, Fig.~\ref{fig:S66_box} depicts a violin plot overlaid a box plot showing the distribution of deviations in separation (upper panel) and energy (lower panel). 
Overall, we find acceptable performance for binding energies 
for all functionals and one can see that all successor functionals indeed improve on the original vdW-DF1 functional to varying degrees. The good performance of our new formulation does carry over from the
S22 data set and we find a median and mean deviation very close to zero. We note that all functionals have outliers, with the binding energy of the neopentane dimer being overestimated for all but the vdW-DF2-B86R, which on the other hand, tends to underestimate binding energies of systems involving aromatic dimers. 
 We also note that the MARD of the S66 set is smaller than that of S22 for all
non-reference system optimized methods,  i.e.\ vdW-DF1, vdW-DF1-cx, vdW-DF2, and vdW-DF2-B86R, as well as for our new modification, with a MARD of 4.62\% compared to 5.72\% for S22. 
For the case of vdW-DF1-optB88, the increased MARD for the S66 set compared to S22 is due to a reduced accuracy for dispersion-bonded systems; 
in particular the binding energies of dimers involving pentane and neopentane, which are not in
the S22 set, are significantly overestimated.
The binding energies of these dimers are also overestimated to a smaller extent with rVV10, but the reduced accuracy
is also due to overestimated binding energies of double-hydrogen bonded dimers.

Looking at the upper panel of Fig.~\ref{fig:S66_box}, we find that
vdW-DF1-optB88 and rVV10 have the most accurate binding separations; however, in essence all but vdW-DF1 and vdW-DF1-cx show good separations, both of which also show a significant spread. Overestimated separations are a well-known shortcoming of  vdW-DF1. The overestimation of vdW-DF1-cx stems from dimers where at least one molecule is an alkane. This result can be related to the fact that this functional was not constructed to provide accurate performance for systems with large reduced gradients $s$.\cite{Berland_2014:exchange_functional}

\subsubsection{The X40 set}

\begin{figure}
\includegraphics[width=\columnwidth]{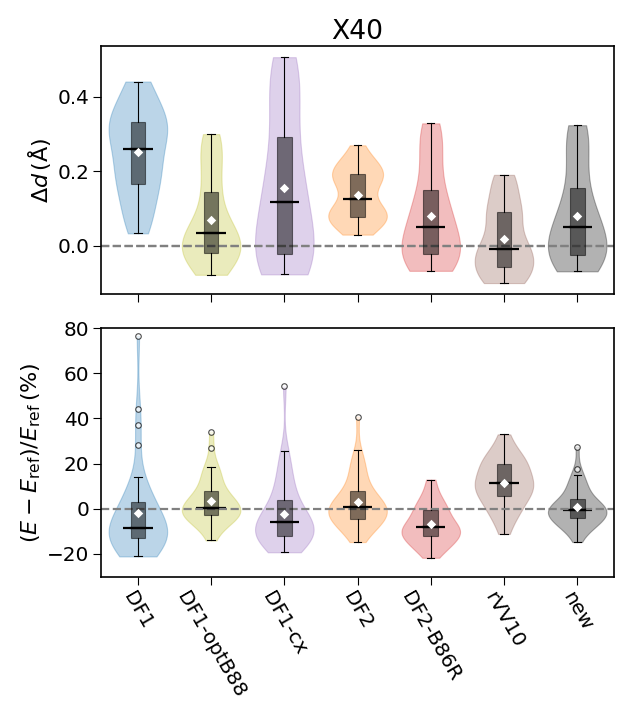}
\caption{\label{fig:X40} Violin/box plots for deviations in separation
and binding energy for the X40 set, see caption of Fig.~\ref{fig:S66_box}
for further details.}
\end{figure}

The X40 set is a set of noncovalently-bonded dimers involving halogenated molecules.
They span a wide variety of different bond characteristics, such as dispersion-dominated F$_2$-methane binding, dipole-dipole bonds, and hydrogen and halide bonds.\cite{Pavel2012:X40dataset}  Figure~\ref{fig:X40} shows a violin/box plot for the X40 set, with statistical data for binding energies summarized in Table~\ref{table:X40}.
The binding separations exhibit similar trends as for S66, but with decreased deviations due to the generally shorter dimer separations. 
Overall, we find that the binding energies of
vdW-DF1-optB88 and our new formulation have the best agreement with the reference data. Like for the S66 data set, our new method follows the same trends as vdW-DF2-B86R but avoids the net underestimation of binding energies exhibited by vdW-DF2-B86R. 
In general, the binding energy MARD is larger for the X40 set than the S66. 
The rVV10 functional has the largest increase in MARD when going from the S66 to X40, from 6.19\% to 15.0\%. In contrast, our new modification merely increases from 4.62\% to 7.06\%.

\begin{figure}
\includegraphics[width=\columnwidth]{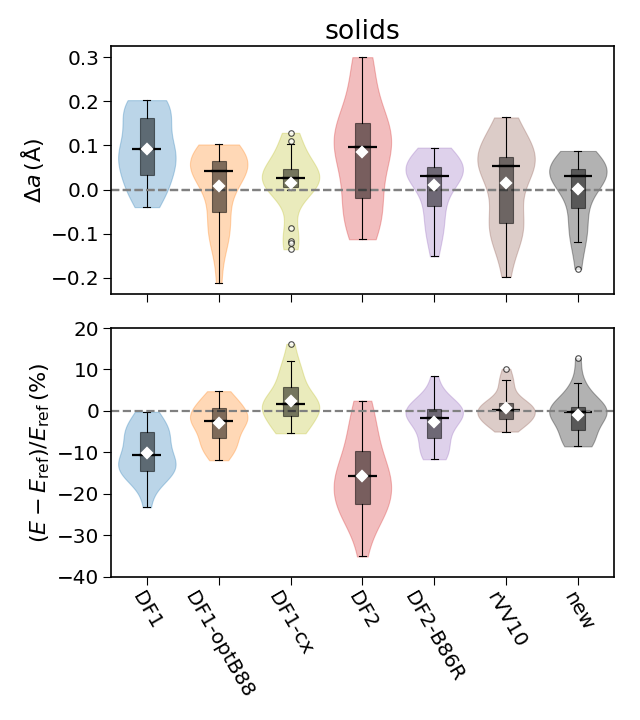}
\caption{\label{fig:solids} Violin/box plots for deviations in lattice constants $a$
and atomization energies of a set of 23 solids, see Table~S7 and S8 for further numerical details and see caption of Fig.~\ref{fig:S66_box}
for a description of violin and box plots.}
\end{figure}

\subsection{Solids}

The original vdW-DF1 and its successor vdW-DF2 both overestimated lattice constants of solids. However, combining vdW-DF correlation with soft exchange functionals with a small ``PBEsol''-type enhancement factor,\cite{PBEsol,
Berland_2014:van_waals} i.e.\ with a Taylor expansion of the form $F_x(s) \approx \mu_{\rm PBEsol}s^2 + \ldots$ , significantly improves solid lattice constants\cite{Klimes_2011:van_waals,Berland_2014:exchange_functional,Berland_2014:van_waals,Berland_2017:asssement_hybrid,yuk2017towards,Ghareee_2017:transition_solids} and in fact, can also improve atomization energies compared to the generalized-gradient approximation.\cite{Klimes_2011:van_waals,Berland_2017:asssement_hybrid}

Figure~\ref{fig:solids} shows results of our performance testing for the same set of 23 solids as considered by Klimes et al.,\cite{Klimes_2011:van_waals} with further numerical data provided in Table~\ref{table:solids} and Tables~S7 and S8 in
the SI.\cite{Note2} The reference data are based on zero-point corrected experimental lattice constants and atomization energies, as detailed in  Ref.~[\onlinecite{Klimes_2011:van_waals}] and references therein. 
All the functionals with soft exchange, i.e vdW-DF1-cx, vdW-DF2-B86R, and our new modifications, give a mean absolute deviation (MAD) for the separations of less than 0.06~\AA,  whereas vdW-DF2 is the least accurate. All functionals, except vdW-DF1 and vdW-DF2, also have an atomization energy MARD smaller than 5\% and our new modification and rVV10 have mean and median deviations close to zero. 
Furthermore, as expected, our modification gives lattice constants that are similar to those of vdW-DF2-B86R, which employs the same exchange functional; atomization energies are also similar but our new formulation shows slightly improved values on average.

\section{Discussion}\label{sec:discussion}

Our study highlights the importance of the specific form of the
wave-vector dependence in $\omega_\bq$ through its switching function
$h(y)$. In doing so we open the door for developments around this so far
unexplored degree of freedom within powerful physical constraints.
Exploiting this freedom trough a judicious choice of the switching
function $h(y)$, we have demonstrated how accurate $C_6$ coefficients
are fully compatible with the constraints of the original van der Waals
density functional, which has been a major criticism of vdW-DF in
general and vdW-DF2 in particular. At the same time, we exploited the
fact that vdW-DF2 predicts trends in the $C_6$ coefficients more
accurately than vdW-DF1.

We do not expect the improvement of $C_6$ coefficients to carry over
to large nanostructures where local-field effects can be important, such
as fullerenes,\cite{hurylula99, Ruzsinszky2012,
Tao_Accurate_fullerene:2016} which would require higher-order terms in
$S$ within vdW-DF. However, at binding separations such higher-order
terms are generally less crucial, at least within vdW-DF,\cite{hybesc14,
TaoHy:2018} as multipole effects are described to second order
in $S$. Moreover, vdW-DF includes non-additive effects
originating from changes in the electronic
density.\cite{Yang_signatures:2018}

Drawing attention to the possibility of adjusting $h(y)$ within vdW-DF
introduces a measure of flexibility that so far has been missing in
standard vdW-DF. In fact, since the release of vdW-DF1 in
2004,\cite{Dion_2004:van_waals} a number of exchange functionals have
been proposed that aim both to improve binding energies and remedy the
notorious overestimation of binding separations present in vdW-DF1 and
to some extent in vdW-DF2. This practice has recently been
criticized\cite{SCAN_VV10:2016} as the large exchange term is fitted to
match the smaller non-local correlation term; colloquially, the ``tail
wags the dog''. Two extreme cases of such a practice are: (i) the
Bayesian-error-estimation-vdW functional (BEEF-vdW),\cite{BEEF-vdW} in
which machine learning was used to determine a semi-local exchange
functional for the vdW-DF2 non-local correlation employing diverse
training sets and (ii) the vdW-DF1-cx
functional,\cite{Berland_2014:exchange_functional} in which the exchange
was designed to be formally as consistent with the vdW-DF1 non-local
correlation as feasible. Because of its flexibility, VV10 (or rVV10) has
seen many adaptions both to different semi-local XC functionals and to
different benchmark sets. \cite{Bjorkmann:solids, SCAN_VV10:2016,
PBsol:VV10, Narbe:2016, semiclassical_VV10:2018} The vdW-DF flexibility
is more limited than that of VV10 both because of the constraints
inherit to the method and the fact that the exchange functional must
still be a good match with the semi-local correlation description of
vdW-DF, which gives a strong preference for ``soft'' exchange functionals.\cite{PBEsol,
Berland_2014:van_waals}


In our new formulation we optimized the parameters of the proposed
$h_{\rm new}(y)$ not only to improve the $C_6$ coefficients but also to
improve the binding energies of the S22 set of molecular dimers. This
improvement carries over to accurate molecular binding properties of the
S66 set and the X40 set. With the choice of B86R, our new formulation
also provides accurate lattice constants for solids. While the performance on molecular dimers is
encouraging, we emphasize that our new proposed switching function is
not necessarily the very best choice for a versatile performance for
different classes of systems as our focus was on improving $C_6$
coefficients and the S22 set of molecular dimers was used to adjust the plasmon model. 
As a simple test on different classes of systems, we also studied adsorption of small molecules in metal organic frameworks and on surfaces. In particular,
for CO$_2$ adsorption in MOF-74 our binding energy shows a 20\% overestimation of experiment compared to 
DF1: 8\%, DF1-optB88: 26\%, DF1-cx: 14\%, DF2: 6\%, DF2-B86R: 4\%, and rVV10: 16\%. 
For the case of benzene on the Au(111) surface, we also find a significant overestimation of binding energy of 22\%, compared to
DF1: $-$20\%, DF1-optb88: 10\%, DF1-cx: 9\%,
DF2: $-$23\%, DF2-b86R: $-$7\%, and rVV10: 15\%. 
Although these are challenging systems, these tests indicate that there is still room for improvement of our new formulation. 
We thus fully expect that new and potentially better forms of
$h(y)$ with significant performance increase over a wide array of
systems will be developed by us or others in the near future and enjoy
widespread usage. For this reason, we do not give our modified
formulation a new name or number within the lineup of vdW-DF1 and
vdW-DF2 functionals.

\section{Summary}\label{sec:summary}

We present a reformulation of the plasmon model that underpins the popular
vdW-DF exchange-correlation functional. Our reformulation is entirely
within the constraint-based framework of the original vdW-DF and
thus inherits its good transferability. Our formulation takes
advantage of some freedom concerning the choice of a switching function
that connects two constrained limits. We use this additional freedom
to correct a long-standing criticism of vdW-DF, i.e.\ the often
wrong $C_6$ coefficients that derive from it. Our work thus proves
that this formal shortcoming is not an inherent flaw of the
vdW-DF formalism, but merely the byproduct of a particular
choice in its parameterization. We test our updated formalism
and find the expected improvement in the $C_6$ coefficients,
but we also find good overall performance with regards to
binding energies and separations for a range
of systems. While we have used this previously unexplored
degree of freedom to improve the $C_6$ coefficients, we see
the main importance of our work in that this freedom
may also be used for further conceptual developments
of vdW-DF and improvements of other properties.

\section*{Acknowledgement}

KB thanks Per Hyldgaard for discussions. Work in Norway was supported by
the Research Council of Norway Project No.\ 250346 and ancillary support
from UiO Energy. All work in the US was entirely supported by NSF Grant
No.\ DMR--1712425.

\appendix
\section{Computational Details}\label{sec:computational_details}

Our new vdW-DF formulation has been implemented in---and all our
calculations have been performed with---the \textsc{quantum espresso}
package.\cite{Giannozzi_2017:advanced_capabilities}
The $C_6$ coefficients were calculated directly from
Eqs.~(\ref{eq:C6}) and (\ref{eq:alpha}). For all our
calculations, we have used the PBE GBRV ultrasoft pseudo potentials due to
their excellent
transferability.\cite{Garrity_2014:pseudopotentials_high-throughput}
However, that database did not include potentials for He, Ne, Ar, and Kr; for those elements
we used the standard PBE RRKJ ultrasoft potentials   
provided by \textsc{quantum espresso}. In all calculations the wave function and density cutoff were $\sim680$~eV (50~Ryd) and $\sim8200$~eV (600~Ryd), respectively. We used an energy convergence criterion 
of $\sim1.36\times 10^{-5}$~eV ($1\times 10^{-6}$~Ryd) for molecular systems and $\sim1.36\times 10^{-7}$~eV ($1\times 10^{-8}$~Ryd) for solids. For metals
and semiconductors a Gaussian smearing was used with a broadening of
$\sim$~0.1~eV (0.00735~Ryd). Lattice parameters and cohesive energies of solids were
determined from the Birch-Murnaghan equation of state.\cite{Birch:1947,Murnaghan:1944} For comparison and to assess the performance of our new formulation, we performed
calculations with the following exchange-correlation functionals: vdW-DF
(also called vdW-DF1 here),\cite{Dion_2004:van_waals}
vdW-DF1-optB88,\cite{Klimes_2010:chemical_accuracy}
vdW-DF1-cx,\cite{Berland_2014:exchange_functional}
vdW-DF2,\cite{Lee_2010:higher-accuracy_van}
vdW-DF2-B86R,\cite{Hamada_2014:van_waals} and
rVV10.\cite{Vydrov_2010:nonlocal_van, Sabatini_2013:nonlocal_van}
The corresponding short names we may use in tables and figures are DF1, DF1-optB88, DF1-cx, DF2, DF2-B86R, and rVV10, respectively.

For the various molecular dimer data sets (S22,\cite{Pavel2006:s22dataset}
S66,\cite{Pavel2011:S66dataset} X40\cite{Pavel2012:X40dataset})
calculations were performed at the
$\Gamma$-point with molecules in a box surrounded by at least 15~\AA\ of
vacuum to minimize spurious interactions with periodic replica.
To test the performance of a given functional on a given data set, we followed
commonly accepted procedures for those data sets. In particular, the monomers
were considered frozen and a fixed number of geometries---representing
different separations---were generated by moving one monomer along an axis 
specified through the optimal structure provided by that data set.
For the S22 data set, a center-of-mass axis was used as suggested by Molner et al.;\cite{Molner2009:s22datset}
for the S66 and X40 data sets an ``interaction coordinate'' was used 
as suggested in the corresponding original works.\cite{Pavel2011:S66dataset, Pavel2012:X40dataset}
Single point calculations were then performed on those geometries to generate binding energy curves,
from which we extracted the binding energy minima and binding separations through fitting
to a Lagrange polynomial near the minimum. 

We have considered 23 solids, semiconductors, and ionic salts as listed in Klimes et al.\cite{Klimes_2010:chemical_accuracy} Periodic solids have been calculated with a $k$-point mesh of $15\times 15\times 15$ for cubic systems and $10\times 10\times 10$ for hexagonal/tetragonal
systems. To calculate the atomization energies, individual atoms have been calculated in a
box with at least $15$~\AA\ of vacuum.

\section{The Internal Functional $\eps^{\rm int}_{\rm xc}$\label{sec:internal}}

In vdW-DF1\cite{Dion_2004:van_waals} the internal functional is given by
the LDA XC energy with Langreth-Vosko exchange gradient corrections for
a slowly varying electron gas,\cite{lavo87} whereas in
vdW-DF2\cite{Lee_2010:higher-accuracy_van} gradient corrections are
given by the large-$N$ asymptote of neutral atoms.\cite{schwinger} In
both cases, it takes the form
\begin{align}
\eps^{\rm int}_{\rm xc}(\br) = \eps^{\rm LDA}_{\rm c}(\br) + \eps_{\rm x}^{\rm LDA}(\br) \big[1 - (Z_{ab}/9) s(\br)^2\big]\,,
\end{align}
with $Z_{ab}= -0.8491$ in vdW-DF1 and $-1.8867$ in vdW-DF2. For our
modification, we employ the same large-$N$ limit for the internal
functional as vdW-DF2 because in our construction it results in more
accurate $C_6$ coefficients than the internal functional of vdW-DF1.

\section{Data}\label{sec:data}

\begin{table*}
    \caption{\label{table:C6}$C_6$ coefficients [Hartree atomic units] for a set of 34 closed-shell atoms and small
    molecules compiled by Vydrov and Voorhis.\cite{Vydrov_2010:dispersion_interactions}
    The $C_6$ coefficients are calculated from
Eqs.~(\ref{eq:C6}) and (\ref{eq:alpha}). For each data set we
    give the mean deviation (MD), mean absolute deviation (MAD),
    mean relative deviation (MRD), and mean absolute relative
    deviation (MARD). The first three data sets
    (vdW-DF1, vdW-DF1-optb88, vdW-DF1-cx) all share the same non-local vdW-DF1
    kernel and only differ in their choice of exchange---thus their results are very close to
    each other and the small differences are a measure of the effect of exchange
    on the $C_6$ coefficients. Similarly, the next two
    data sets (vdW-DF2 and vdW-DF2-B86R) both use the vdW-DF2 kernel (with different exchange)
and produce very comparable results.}

\begin{tabular*}{\textwidth}{@{}l@{\extracolsep{\fill}}rrrrrrrr@{}}\hline\hline
                  &  vdW-DF1  &  vdW-DF1-optB88 &  vdW-DF1-cx &   vdW-DF2 &  vdW-DF2-B86R &     rVV10 &      new &      Ref.\\\hline
He            &     2.98  &            2.94 &        3.09 &      0.76 &          0.78 &      1.45 &     1.82 &     1.46 \\
Ne            &    10.96  &           10.35 &       10.85 &      3.92 &          3.50  &      8.44 &     8.13 &     6.35 \\
Ar            &    77.63  &           74.94 &       75.99 &     30.38 &         29.16 &     70.08 &    67.81 &    64.42 \\
  Kr            &   132.9  &          128.1 &      128.7 &     54.97 &         51.92 &    131.2 &   120.8 &   130.1 \\
  Be            &   304.6  &          296.7 &      310.3 &    106.4 &        108.9 &    186.0 &   253.1 &   214 \\
  Mg            &   723.0  &          671.5 &      727.6 &    246.9 &        243.9 &    425.0 &   568.0 &   627 \\
  Zn            &   271.6  &          244.3 &      256.3 &     79.23 &         78.66 &    163.0 &   183.3 &   284 \\
  H$_2$         &    19.07  &           18.87 &       20.07 &      5.59 &          5.81 &     10.28 &    13.49 &    12.09 \\
N$_2$         &    91.92  &           89.32 &       91.02 &     37.48 &         36.16 &     88.70 &    84.07 &    73.43 \\
  Cl$_2$        &   335.9   &          325.8 &      325.1 &    154.1 &        146.6 &    366.7 &   341.0 &   389.2 \\
  HF            &    27.99  &           27.00 &       27.80 &      9.59 &          9.28 &     21.13 &    21.59 &    19.00 \\
  HCl           &   133.6  &          129.8 &      131.4 &     54.88 &         53.19 &    124.6 &   123.7 &   130.4 \\
  HBr           &   218.4  &          209.9 &      212.5 &     91.89 &         85.78 &    200.2 &   199.6 &   216.6 \\
  CO            &   100.3  &           97.71 &       99.69 &     40.24 &         39.32 &     93.51 &    91.42 &    81.40 \\
  CO$_2$        &   143.1  &          139.6 &      141.0 &     62.52 &         60.39 &    159.4 &   140.4 &   158.7 \\
  CS$_2$        &   661.8  &          643.7 &      645.0 &    310.9 &        299.7 &    739.4 &   697.1  &   871.1 \\
  OCS           &   356.0  &          346.6 &      348.2 &    163.0 &        157.2 &    395.6 &   365.5 &   402.2 \\
  N$_2$O        &   155.7   &          151.4 &      153.1 &      66.93 &         64.86 &    172.4 &   150.8 &   184.9 \\
  CH$_4$        &   138.5  &          136.5 &      141.1 &     56.11 &         56.98 &    129.6 &   132.4 &   129.6 \\
  CCl$_4$       &  1644  &         1600 &     1592  &    828.7 &        792.8 &   2044 &  1844 &  2024 \\
  NH$_3$        &   107.6  &          103.8 &      108.1 &     39.99 &         39.22 &     82.78 &    91.19 &    89.03 \\
  H$_2$O        &    59.24  &           57.08 &       58.94 &     21.21 &         20.52 &     44.95 &    47.72 &    45.29 \\
  SiH$_4$       &   379.2  &          374.6 &      388.5 &    162.2 &        165.7  &    344.6 &   385.0 &   343.9 \\
  SiF$_4$       &   416.2  &          404.7 &      408.1 &    187.0 &        182.1 &    455.8 &   423.5 &   330.2 \\
  H$_2$S        &   217.7  &          211.9 &      215.8 &     90.59 &         89.21 &    200.3 &   207.4 &   216.8 \\
  SO$_2$        &   274.8  &          266.5 &      268.5 &    123.5 &        118.5 &    305.2 &   275.6 &   294.0 \\
  SF$_6$        &   655.3  &          634.9 &      635.3 &    319.5  &        307.9 &    869.9 &   716.3 &   585.8 \\
  C$_2$H$_2$    &   224.6  &          218.5  &      223.3 &     94.02 &         92.22 &    210.3 &   214.4 &   204.1 \\
  C$_2$H$_4$    &   297.4  &          291.0 &      298.1 &    127.4  &        127.0 &    297.3 &   295.1 &   300.2 \\
  C$_2$H$_6$    &   367.4  &          362.3 &      372.0 &    162.1  &        163.9 &    396.6 &   380.9 &   381.8 \\
  CH$_3$OH      &   225.5  &          220.7  &      226.3 &     94.96 &         94.50  &    226.1 &   219.7 &   222.0 \\
  CH$_3$OCH$_3$ &   513.3  &          505.1 &      518.4 &    228.5 &        229.6 &    567.9 &   533.7 &   534.0 \\
  C$_3$H$_6$    &   534.9  &          527.6 &      538.6 &    250.6 &        251.6 &    632.6 &   584.8 &   630.8 \\
  C$_6$H$_6$    &  1411  &         1385 &     1403 &    701.5 &        694.0 &   1838 &  1614  &  1723 \\[1ex]
  MD            & $-$20.23  &        $-$29.80 &    $-$24.05 & $-$203.3 &     $-$206.5 &      2.42 & $-$15.41 & ---      \\  
  MAD           &    51.06  &           52.87 &       53.92 &    203.3  &        206.5 &     36.20 &    38.11 & ---      \\
  MRD [\%]      &     11.36  &            8.02 &         10.91 &    $-$55.64 &        $-$56.58 &   1.24 &  0.97 & ---       \\
  MARD [\%]     &     19.97  &            19.10 &        20.72 &      55.64 &          56.58 &      10.73 &    11.13  & ---      \\\hline\hline
  \end{tabular*}
  \end{table*}

\begin{table*}
\caption{\label{table:S22} Statistical analysis of the performance of various van der Waals functionals for the binding energies of the S22 set, see Table~S1 and S2 for details.}
\begin{tabular*}{\textwidth}{@{}l@{\extracolsep{\fill}}ccccccr@{}}\hline\hline
 &    vdW-DF1 &  vdW-DF1-optB88 &  vdW-DF1-cx &    vdW-DF2 &  vdW-DF2-B86R &  rVV10 &    new\\\hline
{\it Hydrogen-bonded}\\
\quad MD [meV]  &   103 &        14.7 &    34.9 &   58.0 &      25.1 &  $-$21.4 &    14.9 \\
\quad MAD [meV] &   103 &        15.7 &    34.9 &   58.0 &      25.1 &   21.4 &  15.9 \\
\quad MRD [\%]    & $-$17.2 &       $-$3.75 &   $-$7.67 & $-$8.50 &     $-$4.72 &   4.50 & $-$3.30 \\
\quad MARD [\%]    &  17.2 &        3.88 &    7.67 &  8.50 &      4.72 &   4.50 &  3.43 \\[1ex]
{\it Dispersion-bonded}\\
\quad MD [meV]  &  3.98 &       $-$20.4 &    3.95 &  8.17 &     20.5 &   2.98 & $-$9.76 \\
\quad MAD [meV] &  15.8 &        22.4 &    10.9 &  13.0 &     20.5 &   7.28 &  12.7 \\
\quad MRD [\%]     &  9.26 &        9.95 &    3.82 &  2.88 &     $-$11.9 & $-$1.54 &  3.43 \\
\quad MARD [\%]    &  13.5 &        13.0 &   9.66 &  8.26 &      11.9 &   4.73 &   7.98 \\[1ex]
{\it Mixed}\\
\quad MD [meV]  &  22.2 &        4.74 &    16.4 &  21.1 &      23.3 &   6.53 &  10.4 \\
\quad MAD [meV] &  23.0  &        6.70 &    16.5 &  21.9 &      23.3 &   8.29 &  10.4 \\
\quad MRD [\%]     & $-$10.2 &       $-$1.88 &  $-$8.16 & $-$10.4 &     $-$13.2 &  $-$2.63 & $-$5.40 \\
\quad MARD [\%]    &  11.4 &        3.89 &    8.24 &  11.7 &      13.2 &   5.31 &  5.42 \\[1ex]
{\it Full set}\\
\quad MD [meV]  &  41.4 &       $-$1.24 &    17.8 &  28.2 &      22.9 &  $-$3.64 &  4.49 \\
\quad MAD [meV] &    46.0 &        15.3 &    20.3 &  30.2 &      22.9 &   12.1 &    13.0 \\
\quad MRD [\%]     & $-$5.35 &        1.83 &   $-$3.65 & $-$4.97 &       $-$10.0 & 0.04 & $-$1.52 \\
\quad MARD [\%]    &    14.0 &        7.19 &    8.58 &  9.42 &        10.0 &   4.84 &  5.72\\\hline\hline
\end{tabular*}
\end{table*}

\begin{table*}
\caption{\label{table:S66}Statistical analysis of the performance of various van der Waals functionals for the binding energies of the S66 set, see Table~S3 and S4 for details.}
\begin{tabular*}{\textwidth}{@{}l@{\extracolsep{\fill}}ccccccr@{}}\hline\hline
&    vdW-DF1 &  vdW-DF1-optB88 &  vdW-DF1-cx &    vdW-DF2 &  vdW-DF2-B86R &  rVV10 &    new \\\hline
{\it Hydrogen-bonded}\\
\quad MD [meV]  &  51.4 &       $-$1.47 &    14.6 &  19.9 &      6.92 &    $-$25.0 & $-$0.45 \\
\quad MAD [meV] &  51.4 &         8.50 &    17.4 &  20.6 &      13.1 &     25.0 &   9.11 \\
\quad MRD [\%]     & $-$12.4 &       0.39 &   $-$4.34 & $-$3.81 &     $-$2.00 &   7.06 & 0.10 \\
\quad MARD [\%]    &  12.4 &        2.58 &    5.22 &  4.25 &      3.85 &   7.06 &   2.69 \\[1ex]
{\it Dispersion-bonded}\\
\quad MD [meV]  & $-$6.46 &       $-$20.9 &    2.02 &  $-$1.38 &     19.0 &   $-$2.50 & $-$7.66 \\
\quad MAD [meV] &    13.0 &        20.9 &    7.71 &   10.6 &      19.0 &   7.84 &  8.06 \\
\quad MRD [\%]     &  8.11 &        15.6 &   0.77 &   4.22 &     $-$11.7 &   4.08 &  6.13 \\
\quad MARD [\%]    &  10.4 &        15.6 &    5.12 &   7.82 &      11.7 &   6.21 &  6.39 \\[1ex]
{\it Mixed}\\
\quad MD [meV]  &  13.7 &      $-$0.83 &    13.1 &  11.8 &      19.3 &  0.98 &  6.08 \\
\quad MAD [meV] &  16.3 &        6.24 &    13.5 &    14.0 &     19.3 &   7.87 &  7.54 \\
\quad MRD [\%]     & $-$7.70 &       0.63 &   $-$8.10 & $-$6.70 &     $-$12.3 & $-$0.46 & $-$3.79 \\
\quad MARD [\%]    &  9.83 &        4.15 &    8.49 &   8.60 &      12.3 &   5.15 &   4.80 \\[1ex]
{\it Full set}\\
\quad MD [meV]  &  19.8 &       $-$8.05 &    9.75 &   10.0 &      14.9 &  $-$9.29 & $-$0.99 \\
\quad MAD [meV] &  27.4 &        12.1 &    12.9 & 15.1 &      17.0 &   13.8 &   8.27 \\
\quad MRD [\%]     & $-$3.83 &        5.78 &   $-$3.71 & $-$1.90 &     $-$8.51 &   3.74 &   1.02 \\
\quad MARD [\%]    &  10.9 &        7.61 &    6.18 & 6.81 &      9.16 &   6.19 &   4.62\\\hline\hline
\end{tabular*}
\end{table*}

\begin{table*}
\caption{\label{table:X40}Statistical analysis of the performance of various van der Waals functionals for the binding energies of the X40 set, see Table~S5 and S6 for details.}
\begin{tabular*}{\textwidth}{@{}l@{\extracolsep{\fill}}ccccccr@{}}\hline\hline
&    vdW-DF1 &  vdW-DF1-optB88 &  vdW-DF1-cx &    vdW-DF2 &  vdW-DF2-B86R &  rVV10 &  new \\
\hline
MD [meV]  &   13.5 &       $-$6.74 &    2.41 &   2.47 &      3.99 &  $-$18.1 & $-$4.53 \\
MAD [meV] &   19.9 &        10.4 &    14.1 &   9.86 &      11.9 &   19.2 &  9.46 \\
MRD [\%]      & $-$0.12 &        3.93 &  $-$0.95 &   5.11 &      $-$5.93 &   13.4 &  1.86 \\
MARD [\%]     &   15.6 &         7.79 &    12.4 &   10.2 &       9.90 &   15.0 &  7.06 \\
\hline\hline
\end{tabular*}
\end{table*}

\begin{table*}
\caption{\label{table:solids}Statistical analysis of the performance of various van der Waals functionals for the lattice
constants and atomization energies of a set of 23 solids assembled by Klimes et al.\cite{Klimes_2011:van_waals}}
\begin{tabular*}{\textwidth}{@{}l@{\extracolsep{\fill}}ccccccr@{}}\hline\hline
&    vdW-DF1 &  vdW-DF1-optB88 &  vdW-DF1-cx &    vdW-DF2 &  vdW-DF2-B86R &  rVV10 &  new \\\hline
\multicolumn{8}{@{}l}{\it Lattice constants}\\
\quad MD [\AA] &       0.09 &        0.01 &       0.01 &     0.09 &          0.01 &   0.02 & 0.002 \\
\quad MAD [\AA] &    0.10 &        0.07 &       0.06 &     0.12 &          0.06 &   0.09 & 0.06 \\
\quad MRD [\%] &      2.01 &        0.29 &       0.35 &     1.99 &          0.28 &   0.49 & 0.12 \\
\quad MARD [\%] &      2.12 &        1.43 &       1.13 &     2.71 &          1.14 &   1.77 & 1.15 \\[2ex]
\multicolumn{8}{@{}l}{\it Atomization energies}\\
\quad MD [eV] &   $-$0.32 &       $-$0.04 &       0.12 &    $-$0.45 &        $-$0.01 &   0.02 & 0.04 \\
\quad MAD [eV] &    0.32 &        0.10 &       0.16 &     0.47 &         0.10 &   0.08 & 0.11 \\
\quad MRD [\%] &   $-$10.3 &    $-$2.81 &   2.82  &  $-$15.8  &    $-$2.33  & 0.93 & $-$0.52\\
\quad MARD [\%] &    10.3 &        4.31 &       4.87 &    16.2 &         4.35 &   2.99 & 4.08 \\\hline\hline
\end{tabular*}
\end{table*}

\bibliography{references,more_refs}

\end{document}